\documentclass[twocolumn]{aastex62}
\usepackage{CJK}

\usepackage{multirow, amsmath}

\received{2018 August 23}
\revised{2018 September 20}
\accepted{2018 November 12}
\submitjournal{PASP}

\shorttitle{PS1 Point Source Catalog}
\shortauthors{Tachibana \& Miller}

\begin{document}
\begin{CJK*}{UTF8}{gbsn}

\title{A Morphological Classification Model to Identify Unresolved 
       PanSTARRS1 Sources: \\
       Application in the ZTF Real-Time Pipeline
       }

\correspondingauthor{Yutaro Tachibana}
\email{tachibana@hp.phys.titech.ac.jp}

\author[0000-0001-6584-6945]{Yutaro Tachibana}
\affil{Department of Physics, Tokyo Institute of Technology, 2-12-1 Ookayama, Meguro-ku, Tokyo 152-8551,
Japan}
\affil{Department of Physics, Math, and Astronomy, California Institute of Technology, Pasadena, CA, 91125}

\author[0000-0001-9515-478X]{A.~A.~Miller}
\affil{Center for Interdisciplinary Exploration and Research in Astrophysics (CIERA) and Department of Physics and Astronomy, Northwestern University, 2145 Sheridan Road, Evanston, IL 60208, USA}
\affil{The Adler Planetarium, Chicago, IL 60605, USA}



\begin{abstract}

In the era of large photometric surveys, the importance of automated and
accurate classification is rapidly increasing. Specifically, the separation
of resolved and unresolved sources in astronomical imaging is a critical
initial step for a wide array of studies, ranging from Galactic science to
large scale structure and cosmology. Here, we present our method to
construct a large, deep catalog of point sources utilizing Pan-STARRS1 (PS1)
3$\pi$ survey data, which consists of $\sim$3$\times10^9$ sources with
$m\lesssim23.5$\,mag. We develop a supervised machine-learning methodology,
using the random forest (RF) algorithm, to construct the PS1 morphology
model. We train the model using $\sim$5$\times10^4$ PS1 sources with
\textit{HST} COSMOS morphological classifications and assess its performance
using $\sim$4$\times10^6$ sources with Sloan Digital Sky Survey (SDSS)
spectra and $\sim$2$\times10^8$ \textit{Gaia} sources. We construct 11
``white flux'' features, which combine PS1 flux and shape measurements
across 5 filters, to increase the signal-to-noise ratio relative to any
individual filter. The RF model is compared to 3 alternative models,
including the SDSS and PS1 photometric classification models, and we find
that the RF model performs best. By number the PS1 catalog is dominated by
faint sources ($m\gtrsim21$\,mag), and in this regime the RF model
significantly outperforms the SDSS and PS1 models. For time-domain surveys,
identifying unresolved sources is crucial for inferring the Galactic or
extragalactic origin of new transients. We have classified
$\sim$1.5$\times10^9$ sources using the RF model, and these results are used
within the Zwicky Transient Facility real-time pipeline to automatically
reject stellar sources from the extragalactic alert stream.

\end{abstract}




\section{Introduction}\label{sec:intro}

The proliferation of wide-field optical detectors has led to a plethora of
imaging catalogs in the past two decades. Separating unresolved point
sources (i.e., stars and quasi-stellar objects [QSOs]) from photometrically
extended sources (i.e., galaxies) is one of the most challenging and
important steps in the extraction of astronomical information from these
imaging catalogs. For faint sources especially, performing this task well
accelerates our progress in understanding the Universe (e.g.,
\citealt{Sevilla18}). Separating resolved and unresolved sources allows us
to investigate the nature of dark matter by: (i) tracing structure in the
Milky Way halo (e.g., \citealt{Belokurov06}), (ii) measuring galaxy-galaxy
correlation functions (e.g., \citealt{Ross11, Ho15}), and (iii) detecting
the weak lensing signal from cosmic shear \citep{Soumagnac15}. Complete, and
pure, catalogs of galaxies can be used to assess the the geometry of the
Universe \citep{Yasuda01} and the theory of galaxy formation (e.g.,
\citealt{Loveday12, Moorman15}). Finally, for time-domain surveys,
point-source catalogs enable an immediate classification for all newly
discovered variable phenomena as being either Galatic or extragalactic in
origin (e.g., \citealt{Berger12,Miller17}).

Given the many applications for separating point sources from
galaxies, we turn our attention to the Pan-STARRS1 (PS1) 3$\pi$ survey
\citep{Chambers16}, whose $\sim$3$\times 10^{9}$ source catalog provides a
felicitous data set.

The 1.8\,m PS1 telescope is equipped with a wide-field ($\sim$7\,deg$^2$)
1.4 gigapixel camera and is located at Haleakala Observatory in Hawaii
\citep{Hodapp04}. PS1 primarily uses five broadband filters,
$g_{\mathrm{P1}}$, $r_{\mathrm{P1}}$, $i_{\mathrm{P1}}$, $z_{\mathrm{P1}}$,
and $y_{\mathrm{P1}}$ (hereafter $grizy_{\mathrm{P1}}$). The PS1 3$\pi$
survey scanned the entire visible sky ($\delta > -30^\circ$) $\sim$60 times
in the five filters over a 4\,yr time span \citep{Chambers16}. This
repeated imaging was used to create deep stacks \citep{Magnier16a}, with a
typical 5$\sigma$ depth of $\sim$23.2\,mag and a median seeing of
$1.19\arcsec$ in the $r$-band \citep{Tonry12, Schlafly12, Chambers16}. The
first PS1 data release (DR1) provides flux and pixel-based shape
measurements for $>$3 billion sources \citep{Flewelling16}.

Our aim is to develop a large, deep catalog of resolved and unresolved
sources using PS1 data.\footnote{Throughout this paper, we interchangably
use the term star to mean unresolved point source, which includes both stars
and QSOs, while the term galaxy refers to resolved, extended sources.} The
catalog is general purpose and can serve many different science goals,
however, our immediate goal is to support the real-time search for
transients in the Zwicky Transient Facility (ZTF; \citealt{Bellm:18:ZTF}).
Previously, a similar catalog was developed using Palomar Transient Factory
(PTF) data \citep{Miller17}.

The PTF point-source catalog was developed using \texttt{SExtractor}
\citep{bertin96} flux and shape measurements made on deep stacks of PTF
images. Stars and galaxies were separated using a machine learning
methodology built on the random forest (RF) algorithm \citep{Breiman01}.
Briefly, supervised machine learning methods build a non-parametric mapping
between \textit{features}, measured properties, and \textit{labels}, the
target classification, via a training set. The training set contains sources
for which the labels are already known, facilitating the construction of a
features to labels mapping. Following this training, the machine learning
model can produce predictions on new observations where the labels are
unknown.

The PTF point-source catalog was constructed to support the real-time search
for electromagetic counterparts to gravitational wave events. Given that
these events are expected to be very rare (e.g., \citealt{Scolnic18}), the
figure of merit (FoM) for the PTF model was defined as the true positive
rate (corresponding to the fraction of point sources that are correctly
classified) at a fixed false positive rate (fraction of resolved sources that
are misclassified) equal to 0.005 \citep{Miller17}. Maximizing the FoM will
reject as many point sources as possible, while still ensuring that nearly
every extragalactic transient ($\sim$99.5\%) remains in the candidate
stream. While the PTF point-source catalog includes $\sim$1.7$\times 10^8$
objects, the PS1 database includes an order of magnitude more sources.

A resolved--unresolved separation model built on PS1 data will produce
dramatic improvements over the PTF catalog. PS1 observations are deeper,
feature better seeing, and include 5 filters (the PTF catalog was built with
observations in a single filter, $R_\mathrm{PTF}$). Additionally, one of the
12 CCDs in the PTF camera did not work \citep{Law09}, meaning $\sim$8\% of
the $\delta > -30^\circ$ sky has no PTF classifications.

Here, we construct a new morphological classification model using PS1 DR1
data in conjunction with a new machine learning methodology. The model is
trained using \textit{Hubble Space Telescope} observations, which should
provide an improvement over the Sloan Digital Sky Survey (SDSS;
\citealt{York00}) spectroscopic training set used in \citet{Miller17}. As in
\citet{Miller17}, we use the RF algorithm to separate point sources and
extended sources and we optimize our model to maximize the same FoM. Our new
PS1 model outperforms alternatives and has already been incorporated into
the ZTF real-time pipeline. 

Alongside this paper, we have released our open-source analysis, and queries
to recreate the data utilized in this study. These are available online at
\url{https://github.com/adamamiller/PS1_star_galaxy}. The final ZTF--PS1 catalog
created during this study is available as a High Level Science Product via
the Mikulski Archive for Space Telescopes (MAST) at \dataset[doi:10.17909/t9-xjrf-7g34]{http://dx.doi.org/10.17909/t9-xjrf-7g34}.\footnote{\url{https://archive.stsci.edu/prepds/ps1-psc/}}

\section{Model Data}\label{sec:model_data}

Data for the resolved--unresolved model were obtained from the PS1 casjobs
server.\footnote{\url{http://mastweb.stsci.edu/ps1casjobs/home.aspx}} The
PS1 database provides flux measurements via aperture photometry,
point-spread-function (PSF) photometry, and \citet{Kron80}
photometry.\footnote{A subset of bright sources ($i < 21\,\mathrm{mag}$)
outside the Galactic plane have additional photometric measurements, e.g.,
exponential or \citet{Sersic63} profiles, in the \textit{StackModelFitExp}
and \textit{StackModelFitSer} tables, respectively. We ignore these
measurements for this study as they are not available for all sources.}
These flux measurements are produced by PS1 in 3 different ways. The mean
brightness measured on the individual PS1 frames is reported in the
\textit{MeanObject} table, the mean brightness measured via
forced-PSF/aperture photometry on the individual PS1 frames is reported in
the \textit{ForcedMeanObject} table, and finally, the brightness measured on
the full-depth stacked PS1 images is reported in the
\textit{StackObjectThin} table. The \textit{StackObjectAttributes} table
further supplements these tables with point-source object shape
measurements, which prove useful for identifying unresolved sources.
Ultimately, see \S\ref{sec:simple_model}, we use flux measurements from the
\textit{StackObjectThin} table and shape measurements from the
\textit{StackObjectAttributes} table to build our models.

\subsection{The \textit{HST} Training Set} \label{sec:hst_train}

A fundamental challenge in the construction of any supervised machine learning
model is the curation of a high-fidelity training set. A subset of the data
that requires classification must have known labels so the machine can learn
the proper mapping between features and labels. The superior image quality of
the \textit{Hubble Space Telescope} (\textit{HST}) provides exceptionally
accurate morphological classifications, making it an ideal source of a training
set for lower quality ground-based imaging (e.g., \citealt{Lupton01}). The
downside of \textit{HST} is that the field of view is relatively small, so it
is difficult to construct a large and diverse training set suitable for
predictions over the entire sky.

We use the largest contiguous area imaged by \textit{HST}, the 1.64\,deg$^2$
COSMOS field, to construct a training set for our models. Morphological
classifications of \textit{HST} COSMOS sources are provided in
\citet{Leauthaud07}. \citeauthor{Leauthaud07} demonstrate reliable
classifications to $\sim$25\,mag, which is significantly deeper than the
faintest sources detected by PS1. We identify counterparts in the PS1 and
\textit{HST} data by performing a spatial crossmatch between the two catalogs
using a 1\arcsec\ radius.\footnote{This matching radius is the same employed
by PS1 to associate individual detections in the \textit{MeanObject} table
with detections in the \textit{StackObjectAttributes} table.} We further
excluded sources from the \citet{Leauthaud07} catalog with
$\texttt{MAG\_AUTO} > 25$\,mag, as these sources are too faint to be detected
by PS1, meaning their crossmatch counterparts are likely spurious. Following
this procedure, we find that there are 87,431 sources in the
\citet{Leauthaud07} catalog with PS1 counterparts. Of these, 80,974 are
unique in that there is a one-to-one correspondence between \textit{HST}
source and a single PS1 \texttt{ObjID}. The training set is further reduced
to 75,927 once our detection criteria are applied (see
\S\ref{sec:simple_model}), and, of those, only 47,093 have
$\texttt{nDetections} \ge 1$ in the PS1 database (hereafter, the \textit{HST}
training set).\footnote{$\texttt{nDetections}$ refers to the number of
detections in individual PS1 exposures. Thus, \textit{StackObjectThin} souces
can have $\texttt{nDetections}$ = 0 if they are only detected in the PS1
stack images.}

\subsection{The SDSS Training Set}\label{sec:sdss}

\begin{figure*}[htb]
 \centering
  \includegraphics[width=7.2in]{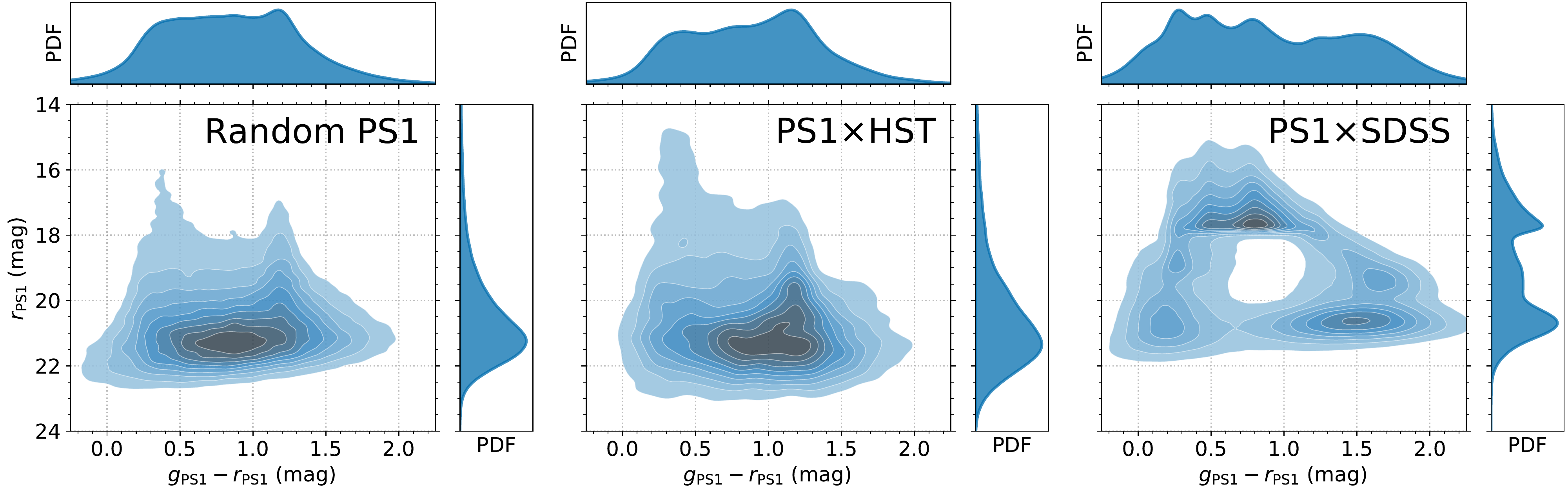}
  \caption{ PS1 color-magnitude diagrams for \textit{left}: $10^6$ randomly
  selected PS1 sources, \textit{center}: the \textit{HST} training set, and
  \textit{right}: the SDSS training set. The primary panels show a
  two-dimensional (2D) Gaussian kernel density estimate (KDE) of the
  probability density function (PDF) of each subset of sources in the
  $r_\mathrm{PS1}$ vs.\ $g_\mathrm{PS1} - r_\mathrm{PS1}$ plane. The shown
  contour levels extend from 0.9 to 0.1 in 0.1 intervals. To the top and
  right of the primary panels are marginalized 1D KDEs of the PDF for the
  $g_\mathrm{PS1} - r_\mathrm{PS1}$ color and $r_\mathrm{PS1}$ brightness,
  respectively. Kron aperture measurements from the \textit{StackObjectThin}
  table are used to estimate each of the PDFs. }
  \label{fig:cmd}
\end{figure*}

The SDSS spectroscopic catalog classifies everything it observes as either a
star, galaxy, or quasi-stellar object (QSO). Using a $1\arcsec$ cross-match
radius, we find 3,834,627 sources with SDSS optical spectra have PS1
counterparts (hereafter, the SDSS training set). Thus, with an orders of
magnitude larger training set, and spectroscopic classifications that should
be both pristine and superior to mophological clsasifications, one might
expect the SDSS training set to be optimal for training the machine learning
model. However, as noted in \citet{Miller17}, the SDSS spectroscopic
targeting algorithms were highly biased, and as a result these sources prove
challenging as a training set.

Color-magnitude diagrams (CMDs) of the \textit{HST} and SDSS training sets
are compared to a random selection of $10^6$ sources from the PS1 database in
Figure~\ref{fig:cmd}. It is clear from Figure~\ref{fig:cmd} that the SDSS
training set is completely different from typical sources in PS1 and that
there are few SDSS sources in the highest density regions of the PS1 CMD.
Given the stark mismatch between typical PS1 sources and the SDSS training
set, we adopt the \textit{HST} training set for the development of our model.
We retain the SDSS training set as an independent test set to assess the
accuracy of the model following construction.

\section{Model Features}\label{sec:model_features}

In addition to developing a training set, we must select features to use as
an input for the model. As noted in \S\ref{sec:model_data}, the PS1 database
provides flux and shape measurements in each of the $grizy_\mathrm{PS1}$
filters. Adopting each of these measurements as features for the model
presents a significant problem: missing data. There are relatively few
sources in the PS1 database that are detected in all 5 filters. Typically,
to cope with missing data one can either (i) remove sources detected in
fewer than 5 filters, or (ii) assign some value, via either imputation
(e.g., \citealt{Miller17}) or the use of a dummy variable, to the missing
data. Given that the vast majority of PS1 sources are faint and are not
detected in all 5 filters, neither of these possiblities is attractive for
our present purposes.

Rather than use the raw features from the database, we engineer a series of ``white flux'' features that combine the relevant measurements across all filters in which a source is detected. In a given filter, a source is detected if the $\mathtt{PSFFlux}_f$, $\mathtt{KronFlux}_f$, 
and $\mathtt{ApFlux}_f$\footnote{These flux measurements are taken from the \textit{StackObjectAttributes} table in the PS1 database. The aperture flux is measured using an ``optimal'' aperture radius based on the local PSF, and corrected based on the wings of the PSF to provide a total flux (for point sources). The Kron flux is measured inside 2.5 times the first radial moment, which is expected to miss up to $\sim$10\% of the total light from galaxies \citep{Magnier16b}.} 
are \textit{all $> 0$}, where the $f$ subscript refers to a specific filter. The ``white flux'' feature is then created as:
\begin{equation}
    \mathtt{white[Feat]} =  \frac{\sum_f^{f = grizy_\mathrm{PS1}} w_f  \, \mathtt{Feat}_f \, \mathrm{det}_f}{\sum_f^{f = grizy_\mathrm{PS1}} w_f}, 
\end{equation}
where the sum is over the 5 PS1 filters, \texttt{Feat} is the feature from the \textit{StackObjectAttributes} table, $\mathrm{det}_f = 1$ if the source is detected in the $f$ filter, as defined above, or $\mathrm{det}_f = 0$ if not detected, and $w_f$ is the weight assigned to each filter:
\begin{equation}
    w_f = \left(\frac{\mathtt{KronFlux}_f}{\mathtt{KronFluxErr}_f}\right)^2,
\end{equation}
equivalent to signal-to-noise ratio (SNR) squared in the given filter. 
Ultimately, the ``white flux'' features correspond to a weighted mean, with weights equal to the square of the SNR \citep{Bevington03}. 

Our final model includes 11 ``white flux'' features to separate resolved and unresolved sources. The database features include: \texttt{PSFFlux},\footnote{For the
\texttt{PSFFlux} feature $w_f =
(\mathtt{PSFFlux}_f/\mathtt{PSFFluxErr}_f)^2$.} \texttt{KronFlux},
\texttt{ApFlux},\footnote{For the \texttt{ApFlux} feature $w_f =
(\mathtt{PSFFlux}_f/\mathtt{PSFFluxErr}_f)^2$.} 
\texttt{ExtNSigma},
\texttt{KronRad}, \texttt{psfChiSq}, \texttt{psfLikelihood},
\texttt{momentYY}, \texttt{momentXY}, \texttt{momentXX}, and
\texttt{momentRH}.\footnote{Prior to their ``white flux'' calculation the
shape features (\texttt{KronRad}, \texttt{momentYY}, \texttt{momentXY},
\texttt{momentXX}, and \texttt{momentRH}) are normalized by the seeing in the
respective bandpass, which we define as the \texttt{psfMajorFWHM} and
\texttt{psfMinorFWHM} added in quadrature. \texttt{KronRad} has units of
arcsec, \texttt{momentRH} has units of arcsec$^{0.5}$, and the remaining
shape features have units of arcsec$^{2}$. They are each normalized by
dividing by the seeing raised to the appropriate power. } The remaining
features in the database were either uninformative or would bias the model,
such as R.A.\ and Dec.\ (see e.g., \citealt{Richards12a}). We do not directly
include \texttt{whitePSFFlux}, \texttt{whiteKronFlux}, and
\texttt{whiteApFlux} in the model. We found that the inclusion of these
features resulted in a bias whereby all sources brighter than $\sim$16\,mag
were automatically classified as point sources. Instead, we include the ratio of the
different flux measures: \texttt{whitePSFKronRatio} =
\texttt{whitePSFFlux}/\texttt{whiteKronFlux}, \texttt{whitePSFApRatio} =
\texttt{whitePSFFlux}/\texttt{whiteApFlux}, as well as a third feature
\texttt{whitePSFKronDist} (see \S\ref{sec:simple_model}).

As we previously alluded to, the primary benefit of the ``white flux''
features is that they can be calculated for every source in PS1 thus
allowing each to be compared on common ground. Furthermore, the SNR for the
``white flux'' features is greater than the SNR for the equivalent feature
in a single filter. The downside of these features is that for some sources,
especially at the bright end, color information is lost. While a blue source
and red source with identical \texttt{whitePSFFlux} values are intrinsically
very different, the ``white flux'' features obscure that information for the
classifier. Ultimately, we tested models using the ``white flux'' features
with and without additional color features and found that they are
statistically equivalent when tested with the \textit{HST} training set.

The direct use of color information as model features would require
reddening corrections for all PS1 sources. Not only is this a daunting task,
but accurate corrections would require a priori knowledge as to which
sources are Galactic and which are extragalactic (e.g., \citealt{Green15}).
The PS1 catalog is being developed precisely to answer this question.
Furthermore, the pencil beam sample from the \textit{HST} training set
traces a narrow range of dust columns, so the application of a model
including color information without reddening corrections would lead to
biased classifications \citep{Sevilla18}, particularly in regions of high
reddening (e.g., the Galactic plane). We conclude that the benefits of the
``white flux'' features, which eliminate the need for reddening corrections,
outweigh any losses from the exclusion of color information.

\begin{figure*}[htb]
 \centering
  \includegraphics[width=5.75in]{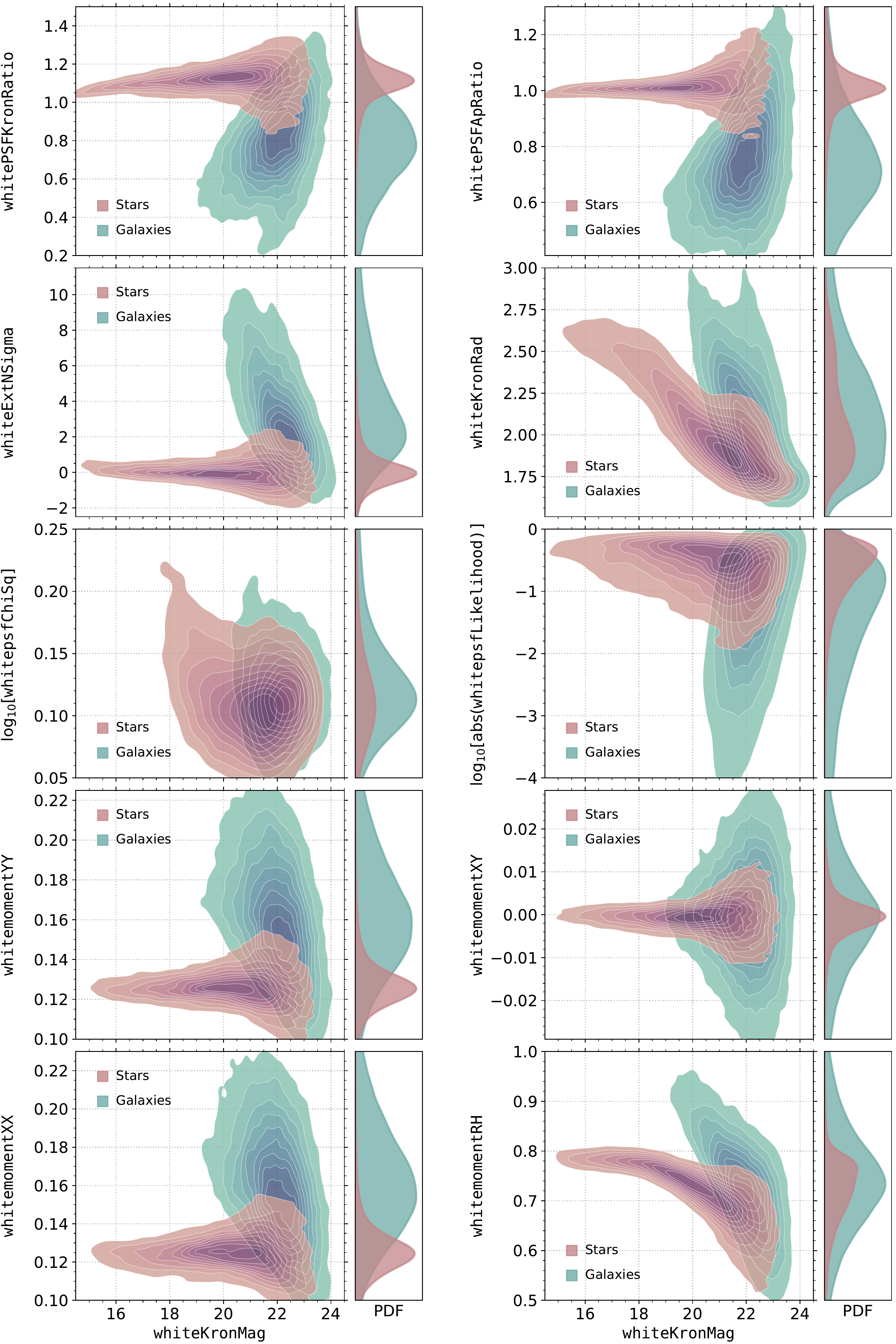}
  \caption{
  The primary square panels show Gaussian KDEs of the PDF for each of the
  ``white flux'' features as a function of \texttt{whiteKronMag}
  ($=-2.5\log_{10}[\mathtt{whiteKronFlux}/3631]$) for all sources in the
  \textit{HST} training set. Unresolved point sources (labeled stars) are
  shown via the red-purple contours, while resolved, extended objects
  (labeled galaxies) are shown via blue-green contours. The shown contour
  levels extend from 0.9 to 0.1 in 0.1 intervals. To the right of each
  primary panel is a marginalized 1D KDE of the PDF for the individual
  features, where the amplitudes of the KDEs have been normalized by the
  relative number of point sources and extended objects. The clear overlap
  between faint resolved and unresolved sources suggests that a machine
  learning model may provide significant improvement over the PS1 and simple
  models.}
  \label{fig:features}
\end{figure*}

The distribution of ``white flux'' features for point sources and extended objects in the
\textit{HST} training set is shown in Figure~\ref{fig:features}
(\texttt{whitePSFKronDist} is shown in Figure~\ref{fig:psfkrondist}). As
might be expected, it is clear from Figure~\ref{fig:features} that point
sources and extended objects are easily separated at the bright end
($\lesssim 20$\,mag), but there is significant overlap in the featurespace
between the two populations at the faint end ($\sim$23\,mag). Machine
learning algorithms are capable of capturing non-linear behavior in
multidimentional data sets, which will prove especially useful for the
sources under consideration in the PS1 data set.

\section{Model Construction}
\subsection{The PS1 Baseline Model}\label{sec:ps1_model}

To establish a baseline for the performance of our resolved--unresolved
separation models we adopt the classification criteria in the PS1
documentation, namely sources with 
$$ \mathtt{iPSFMag} - \mathtt{iKronMag}
> 0.05\;\mathrm{mag},$$ 
are classified as
galaxies.\footnote{\url{https://outerspace.stsci.edu/display/PANSTARRS/How+to
 +separate+stars+and+galaxies}} The documentation notes that this
classification can be performed using photometry from any of the
\textit{MeanObject}, \textit{ForcedMeanObject}, or \textit{StackObjectThin}
tables. The PS1 documentation further notes that this basic cut does not
perform well for sources with $i \gtrsim 21\,\mathrm{mag}$, which
constitutes the majority of sources detected by PS1, and motivates us to
develop alternative models. We use the performance of the $\mathtt{iPSFMag}
- \mathtt{iKronMag} > 0.05\;\mathrm{mag}$ model (hereafter, the PS1 model)
as a baseline to compare to the models discussed below.

\subsection{Simple Model}\label{sec:simple_model}

While our ultimate goal is to build a machine learning model to identify
point sources (\S\ref{sec:rf_model}), we first construct a straightforward
model. This model is inspired by the SDSS \texttt{photo} pipeline
\citep{Lupton01}, and combines the flux in each of the 5 PS1 filters to
improve the SNR relative to any individual band. In addition to being easy
to interpret, this model (hereafter, the simple model), which utilizes the
difference between the PSF flux and the Kron-aperture flux for
classification, serves as an additional baseline to test the need for a more
complicated machine learning model.

It stands to reason that a model built on all five PS1 filters should
outperform a model constructed from a single filter. To that end, we examine
the \texttt{whitePSFKronRatio} (equivalent to $\mathtt{whitePSFMag} -
\mathtt{whiteKronMag}$) to discriminate between resolved and unresolved
sources. The upper left panel of Figure~\ref{fig:features} shows that
sources with $\mathtt{whitePSFKronRatio} \gtrsim 1$ are very likely point sources. A single hard cut on \texttt{whitePSFKronRatio}, similar to the SDSS
\texttt{photo} pipeline or the PS1 model, removes any sense of confidence in
the corresponding classification. For example, a source with
$\mathtt{whitePSFKronRatio} = 1.1$ and $\mathtt{whiteKronMag} \approx
17\,\mathrm{mag}$ is far more likely to be a point source than a source with the
same \texttt{whitePSFKronRatio} value but $\mathtt{whiteKronMag} \approx
23\,\mathrm{mag}$ (see Figure~\ref{fig:features}).

To address this issue of classification confidence, we measure the
orthogonal distance from a line ($\mathtt{whitePSFFlux} = a\times
\mathtt{whiteKronFlux}$) for all sources in the
$\mathtt{whitePSFFlux}$--$\mathtt{whiteKronFlux}$ plane to define the simple model:
\begin{multline}
 \mathtt{whitePSFKronDist}(a) = \\
 \frac{\mathtt{whitePSFFlux} - a\times\mathtt{whiteKronFlux}}{ \sqrt{1 + a^2}},
 \label{eqn:psfkrondist}
\end{multline}
where $a$ is the slope of the line. For $a \approx 1$, which is similar to a
hard cut with $\mathtt{whitePSFKronRatio} = a$, bright point sources will
have large, positive values of \texttt{whitePSFKronDist}, while bright
extended objects will have large, negative values of
\texttt{whitePSFKronDist}. Simultaneously, faint sources, which are more
difficult to classify owing to the lower SNR, will have small values of
\texttt{whitePSFKronDist}. The simple model allows us to produce a rank
ordered classification, which in turn allows us to evaluate the optimal
classification threshold for the separation of resolved and unresolved 
sources (see \S\ref{sec:comp_hst}).

The optimal value for $a$ is determined via $k$-fold cross validation
(CV).\footnote{In $k$-fold CV, $1/k$ of the training set is withheld during
model construction, and the remaining $1-1/k$ fraction of the training set is
used to predict the classification of the withheld data. This
procedure is repeated $k$ times, with every training set source
being withheld exactly once, so that predictions are made for each source in
the training set enabling a measurement of the FoM.} We adopt identical
procedures to optimize both the simple model and the machine learning model
(see \S\ref{sec:rf_model}). We employ the use of an inner and outter CV
loop, both of which have $k = 10$ folds. In the outter CV loop, the training
set is split into 10 separate partitions, each iteratively withheld from the
training. For each partition in the outter CV loop, an inner 10-fold CV is
applied to the remaining $\sim$90\% of the training set to determine the
optimal model parameters. Predictions on the sources withheld in the outter
loop are made with the optimal model from the inner loop to provide model
predictions for every source in the training set. We adopt final, optimal
tuning parameters from the mean of the values determined in the inner CV.

For the simple model, we employ a grid search over $a$ in the inner CV loops
to maximize the FoM and thereby determine the optimal value of $a$.
Initially, a wide grid from 0 to 2 was searched, followed by a fine grid
search over $a$ from 0.75 to 1.25 with step size = 0.0025. The average
optimal $a$ from the inner loops, and hence final model value, is 0.91375,
with sample standard deviation $\sim$0.01. From this procedure, we find that
for the simple model the $\mathrm{FoM} = 0.62 \pm 0.02$, where the
uncertainty is estimated from the scatter in the outter CV
folds.\footnote{We find that PS1 flux measurements from the
\textit{StackObjectThin} table produce a higher FoM for the simple model
than flux measurements from the \textit{MeanObject} and
\textit{ForcedMeanObject} tables. Thus, we adopt \textit{StackObjectThin}
fluxes for both the simple model and the machine learning model, as noted in
\S\ref{sec:model_data}.}

\begin{figure}[t]
 \centering
  \includegraphics[width=3.35in]{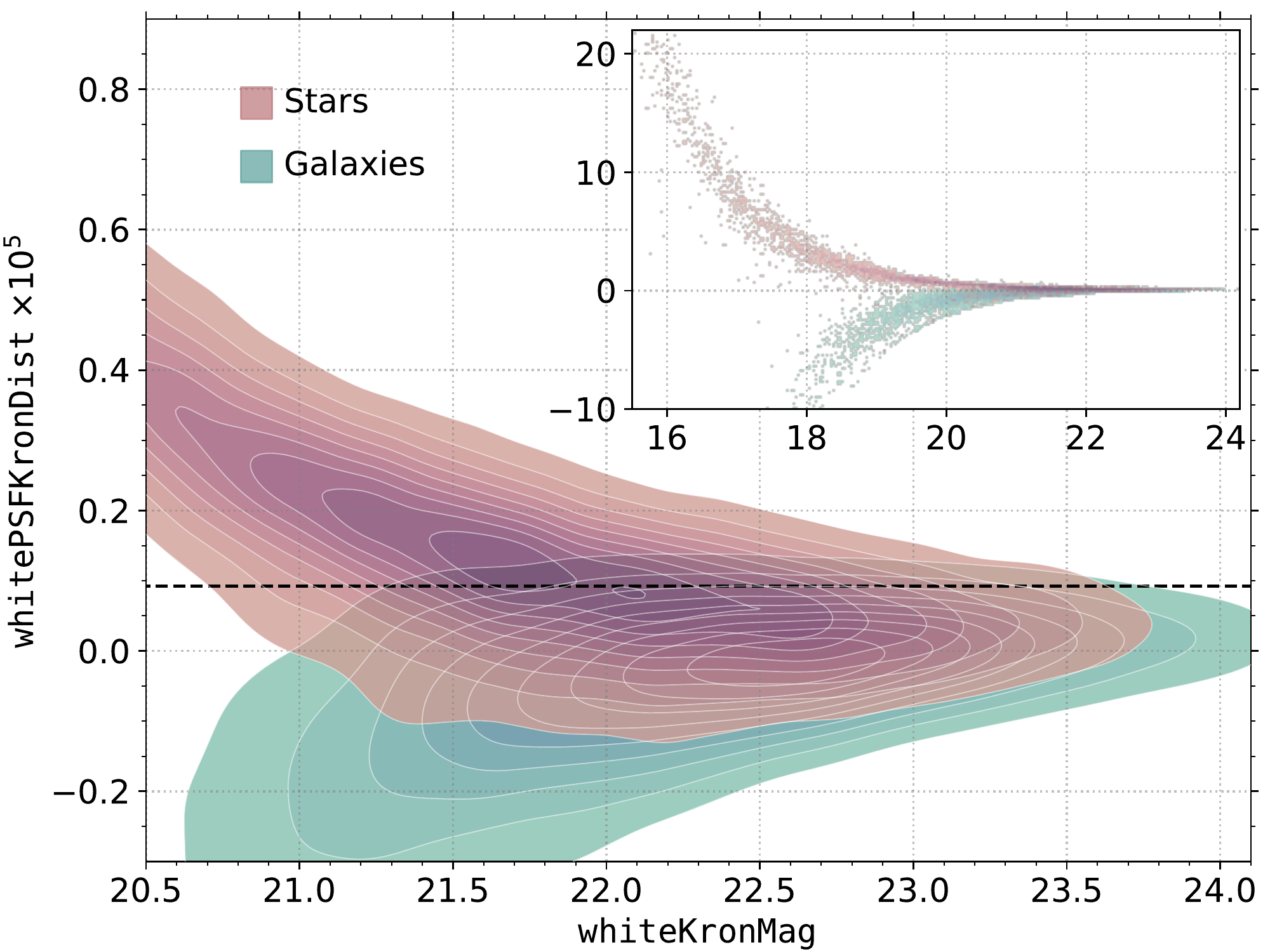}
  \caption{ The distribution of $\mathtt{whitePSFKronDist}$ values for
  \textit{HST} training set point sources (labeled stars) and extended
  objects (labeled galaxies) as a function of \texttt{whiteKronMag}. The
  colors and contours are the same as Figure~\ref{fig:features}. The
  horizontal dashed line shows the optimal threshold
  ($\mathtt{whitePSFKronDist} \ge 9.2 \times 10^{-7}$) for
  resolved--unresolved classification. The upper-right inset shows a
  zoom-out highlighting the stark difference between stars and galaxies at
  the bright end. }
  \label{fig:psfkrondist}
\end{figure}

The distribution of $\mathtt{whitePSFKronDist}(a=0.91375)$ is shown for the
\textit{HST} training set in Figure~\ref{fig:psfkrondist}.
$\mathtt{whitePSFKronDist}$ provides an excellent discriminant between
bright ($\lesssim 20\,\mathrm{mag}$) point sources and extended objects. We
further find that adopting a point source classification threshold of
$\mathtt{whitePSFKronDist} \ge 9.2 \times 10^{-7}$ produces a classification
accuracy of $\sim$91\%.

\subsection{Random Forest Model}\label{sec:rf_model}

\subsubsection{The Random Forest Algorithm}\label{sec:rf_alg}

Based on its success in previous astronomical applications (e.g.,
\citealt{Richards12a, Huppenkothen17, Brink13, Wright15, Goldstein15}),
including morphological classification (e.g., \citealt{Vasconcellos11,Miller17}),
we adopt the RF algorithm \citep{Breiman01} for our machine learning model.
In fact, following the comparison of several different algorithms it was
recently found that ensemble tree-based methods, such as RF, perform best
when separating stars and galaxies \citep{Sevilla18}. In future work, we
will consider alternative ensemble methods (such as adaptive boosting;
\citealt{Freund97}), which is found to slightly outperform RF for
similar problems \citep{Sevilla18}.

Briefly, RF is built on decision tree models \citep{Quinlan93} that utilize
bagging \citep{Breiman96}, wherein bootstrap samples of the training set are
used to train each of the $N_{\mathrm{tree}}$ individual trees. Within the
individual trees, only $m_{\mathrm{try}}$ randomly selected features are
used to separate sources at each node, and nodes cannot be further split if
there are fewer than \texttt{nodesize} sources in the node. The randomness
introduced by both bagging and the use of $m_{\mathrm{try}}$ features
reduces the variance of RF predicitions relative to single decision tree
models. Final RF classifications are determined via a majority vote from
each of the $N_{\mathrm{tree}}$ individual trees. Thus, RF models are
capable of producing low-variance, low-bias predictions. We utilize the
\texttt{Python scikit-learn} implementation of the RF algorithm
\citep{Pedregosa12} in this study.

\subsubsection{Feature Selection}

While the RF algorithm is relatively insensitive to correlated and/or
weak/uninformative features (e.g., \citealt{Richards12a}), we nevertheless
investigate if removing features from our feature set improves the model
performance.\footnote{For example, we find a strong correlation between
\texttt{whiteExtNSigma} and \texttt{whitepsfLikelihood} (Pearson correlation
coefficient $r = 0.85$), which can potentially lead to overfitting.} We do
this via forward and backward feature selection \citep{Guyon03}. Forward and
backward feature selection involve the iterative addition or removal of
features from the model, respectively. Like \citet{Richards12a}, we rank
order the features for either addition or subtraction based on their
RF-determined importance \citep{Breiman02}. This method shows
\texttt{whitePSFKronDist} to be the most important feature, and we find that
removing features does not improve the CV FoM. We therefore include all 11
``white flux'' features from \S\ref{sec:model_features} in the final RF
model.

\subsubsection{Optimizing the Model Tuning Parameters}

As noted in \S\ref{sec:simple_model}, we optimize the RF model tuning
parameters via an outter and inner 10-fold CV procedure. We perform a grid
search over $N_{\mathrm{tree}}$, $m_{\mathrm{try}}$ and \texttt{nodesize},
and find that the FoM for the \textit{HST} training set is maximized with
$N_{\mathrm{tree}} = 400$, $m_{\mathrm{try}} = 4$, and $\mathtt{nodesize} =
2$. The final model FoM is not strongly sensitive to the choice of these
parameters: changing any of the optimal parameters by a factor of $\sim$2
does not decrease the optimal CV FoM, $\sim$0.71, by more than the scatter
measured from the individual folds, $\sim$0.02. Finally, while a detailed
comparison is presented in \S\ref{sec:comp_hst}, we note that the RF model
significantly outperforms the simple model based on the CV FoM.

\section{Classification Performance}

\subsection{PS1, Simple, and RF Model Comparison}\label{sec:comp_hst}

We assess the relative performance of the RF model by comparing it to both
the PS1 and simple models. To do so, we select the subset of sources from
the \textit{HST} training set that have $\mathrm{det}_{i_\mathrm{PS1}} = 1$
(the PS1 model cannot classify sources that are not detected in the
$i_\mathrm{PS1}$ band), which results in 40,098 sources.

\begin{figure}[t]
 \centering
  \includegraphics[width=3.35in]{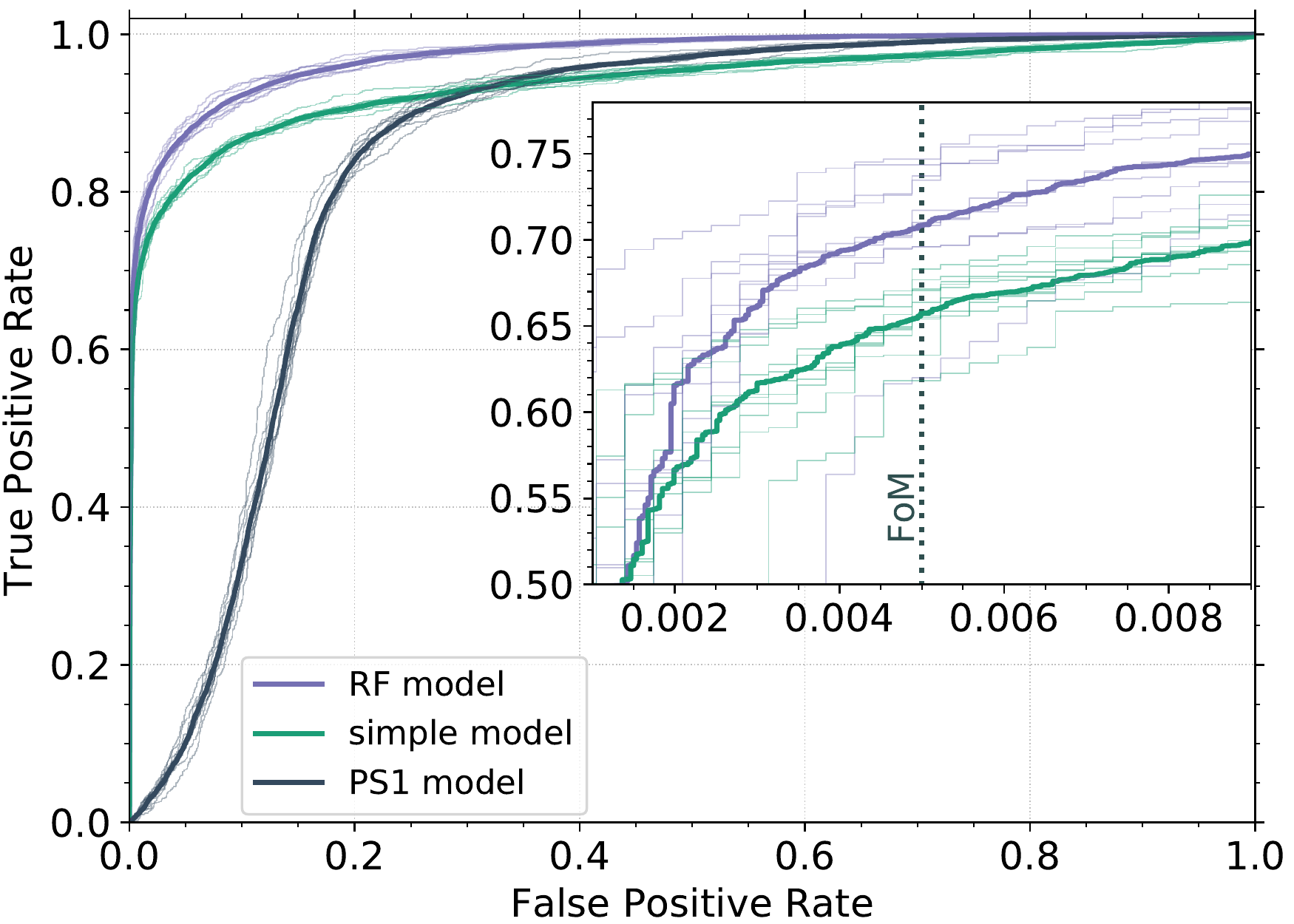}
  \caption{ ROC curves comparing the relative performance of the PS1,
  simple, and RF models as tested by the subset of \textit{HST} training set
  sources with $i_\mathrm{PS1}$ detections. The thick, solid slate gray, green,
  and purple lines show the ROC curves for the PS1, simple, and RF models,
  respectively. The light, thin lines show the ROC curves for the individual
  CV folds. The inset on the right shows a zoom in around FPR = 0.005, shown
  as a dotted vertical line, corresponding to the FoM (the PS1 model is not
  shown in the inset, because it has very low FoM). }
  \label{fig:cvroc_hst}
\end{figure}

Figure~\ref{fig:cvroc_hst} shows that the RF model and simple model provide
substantial improvements over the PS1 model, with $\sim$10,000\% and
$\sim$9,200\% respective increases in the FoM relative to the PS1 model. We
additionally show Receiver Operating Characteristic (ROC) curves for the 3
models in Figure~\ref{fig:cvroc_hst}. ROC curves show how the true positive
rate (TPR)\footnote{$\mathrm{TPR} = \mathrm{TP}/(\mathrm{TP} +
\mathrm{FP})$, where TP is the total number of true positive classifications
and FP is the number of false positives.} and false positive rate
(FPR)\footnote{$\mathrm{FPR} = \mathrm{FP}/(\mathrm{FP}+\mathrm{TN})$, where
TN is the number of true negatives.} vary as a function of classification
threshold. As a reminder, point sources are considered the positive class in
this study. To construct ROC curves for the simple and PS1 models we vary
the classification thresholds from $\mathtt{whitePSFKronDist} = 4.24\times
10^{-3}$ to $-16.23\times10^{-3}$ and $\mathtt{iPSFMag} - \mathtt{iKronMag}
= 5.10\;\mathrm{mag}$ to $-2.81\;\mathrm{mag}$, respectively.
Figure~\ref{fig:cvroc_hst} highlights the strength of the simple model
approach: by using a metric that essentially captures both the difference
between the PSF and Kron flux measurements \textit{and} the SNR, the simple
model produces much higher TPR at low FPR than the PS1 model, which does not
capture information about the SNR.

Summary statistics showing the superior performance of the RF model relative
to the simple and PS1 models are presented in Table~\ref{tbl:hst_cv}. These
statistics include the FoM, the overall classification accuracy, and the
integrated area under the ROC curve (ROC AUC) of the 3 models as evaluted on
the subset of \textit{HST} training set sources with $i_\mathrm{PS1}$
detections. We use 10-fold CV to measure the summary statistics, with
identical folds for each model. Strictly speaking, this CV procedure is only
needed for the RF model, which needs to be re-trained for every fold, but
testing the simple and PS1 models on the individual folds provides an
estimate in the scatter of the final reported metrics. From
Table~\ref{tbl:hst_cv} it is clear that the RF model greatly outperforms the
simple and PS1 models.

\begin{deluxetable}{cccc}
    \tablecolumns{4} 
    \tablewidth{0pt} 
    \tablecaption{ CV Results for the \textit{HST} Training Set \label{tbl:hst_cv}}
    \tablehead{ 
    \colhead{model} & \colhead{FoM} & \colhead{Accuracy} & \colhead{ROC AUC}
    }
    \startdata
    RF & {\bf 0.707} $\pm$ 0.036 & {\bf 0.932} $\pm$ 0.003 & {\bf 0.973} $\pm$ 0.002 \\
    simple & 0.657 $\pm$ 0.020 & 0.916 $\pm$ 0.003 & 0.937 $\pm$ 0.004 \\
    PS1 & 0.007 $\pm$ 0.003 & 0.810 $\pm$ 0.006 & 0.851 $\pm$ 0.006 \\
    \enddata
    \tablecomments{Uncertainties represent the sample standard deviation for the 10 individual folds used in CV. For each metric, the model with the best performance is shown in bold.}
\end{deluxetable}

\begin{figure}[t]
 \centering
  \includegraphics[width=3.35in]{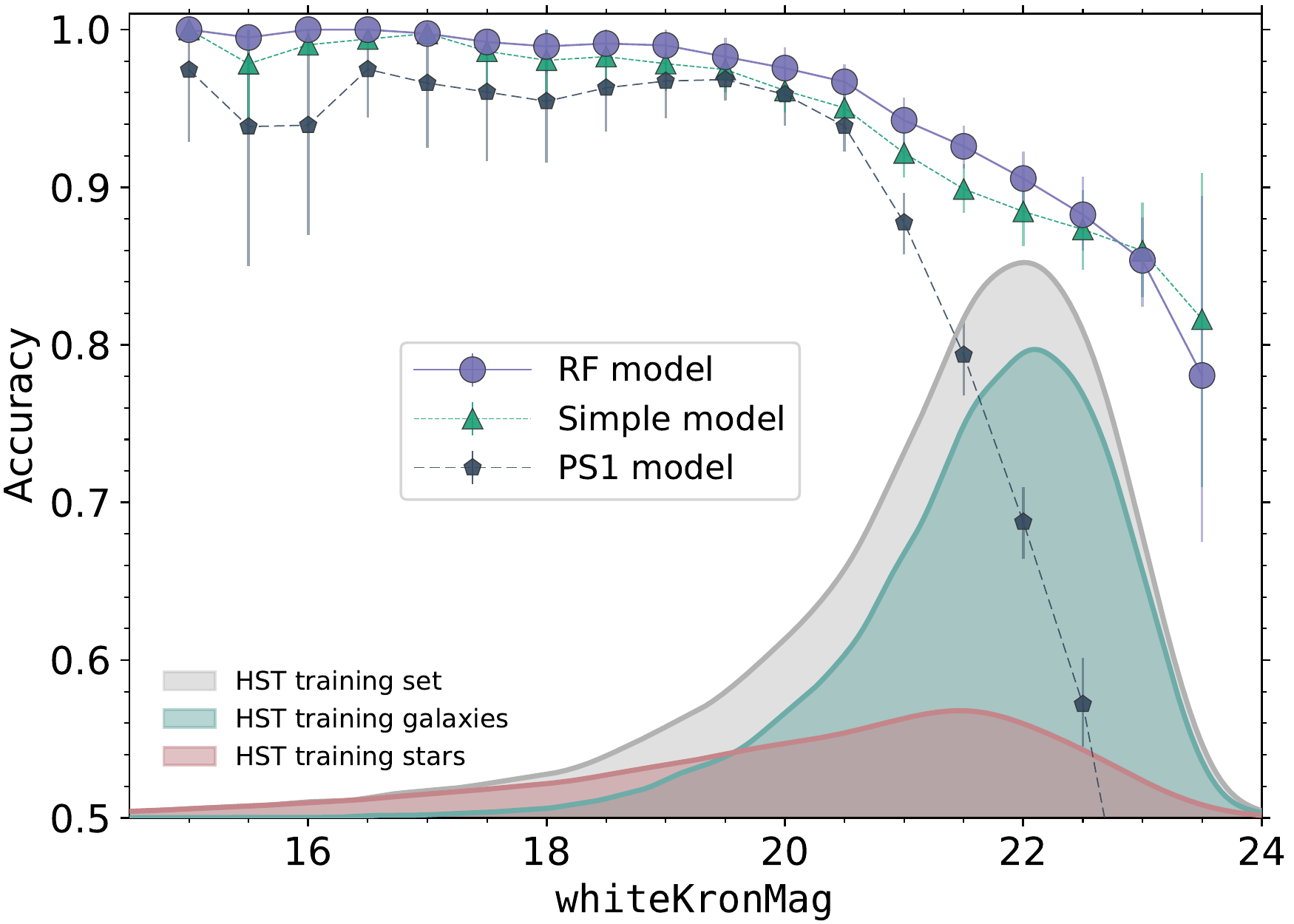}
  \caption{Model accuracy as a function of \texttt{whiteKronMag} evaluated
  on the subset of \textit{HST} training set sources with $i_\mathrm{PS1}$
  detections. Accuracy curves for the PS1, simple and RF models are shown as
  slate gray pentagons, green triangles, and purple circles, respectively.
  The bin widths are 0.5\,mag, and the error bars represent the 68\%
  interval from bootstrap resampling. Additionally, a Gaussian KDE of the
  PDF for the $i_\mathrm{PS1}$-detection subset of \textit{HST} training
  set, as well as the point sources (labeled stars) and extended objects
  (labeled galaxies) in the same subset is shown in the shaded gray, red,
  and green regions, respectively. The amplitude of the star and galaxy PDFs
  have been normalized by their relative ratio compared to the full
  $i_\mathrm{PS1}$-band subset.} 
  \label{fig:cvacc_hst}
\end{figure}

The classification accuracy for each model as a function of
\texttt{whiteKronMag} is shown in 0.5\,mag bins in
Figure~\ref{fig:cvacc_hst}. The accuracies are estimated via 10-fold CV (see
above) and the uncertainties represent the inter-68\% interval from
100 bootstrap samples within each bin. 
The classification thresholds for the RF, simple, and PS1 models are 0.5,
$9.2 \times 10^{-7}$, and 0.05, respectively. Again, the RF and simple
models provide a significant improvement over the PS1 model. The PS1 model
provides classification accuracies $\gtrsim$90\% for sources with
$\mathtt{whiteKronMag} \lesssim 21\,\mathrm{mag}$, but precipitously
declines for fainter sources. The RF and simple models have similar curves
with the RF model performing slightly better, as is to be expected given
that the RF model uses 10 additional features beyond
\texttt{whitePSFKronDist}.

\subsection{Model Evaluation via an Independent Test Set}

\begin{deluxetable}{cccc}
    \tablecolumns{4} 
    \tablewidth{0pt} 
    \tablecaption{SDSS Test Set Metrics\label{tbl:sdss_per}}
    \tablehead{ 
    \colhead{model} & \colhead{FoM} & \colhead{Accuracy\tablenotemark{a}} & \colhead{ROC AUC}
    }
    \startdata
    RF & \textbf{0.843} $\pm$ 0.001 & 0.9625 $\pm$ 0.0001 & \textbf{0.98713} $\pm$ 0.00007 \\
    simple & 0.798 $\pm$ 0.002  & 0.9557 $\pm$ 0.0001 & 0.98503 $\pm$ 0.00008 \\
    PS1 & 0.290 $\pm$ 0.004 & 0.9612 $\pm$ 0.0001 & 0.98411 $\pm$ 0.00007 \\
    SDSS & 0.777 $\pm$ 0.003 & \textbf{0.9713} $\pm$ 0.0001 & 0.98660 $\pm$ 0.00008 \\
    \enddata
    \tablecomments{Uncertainties represent the sample standard deviation for 100 bootstrap samples of the SDSS test set. For each metric, the model with the best performance is shown in bold.} 
    \tablenotetext{a}{Classification accuracies are evaluated using classification cuts of $0.5$, $9.2 \times 10^{-7}$, $0.05$, and $0.145$ for the RF, simple, PS1, and SDSS models, respectively.}
\end{deluxetable}

\begin{figure}[t]
 \centering
  \includegraphics[width=3.35in]{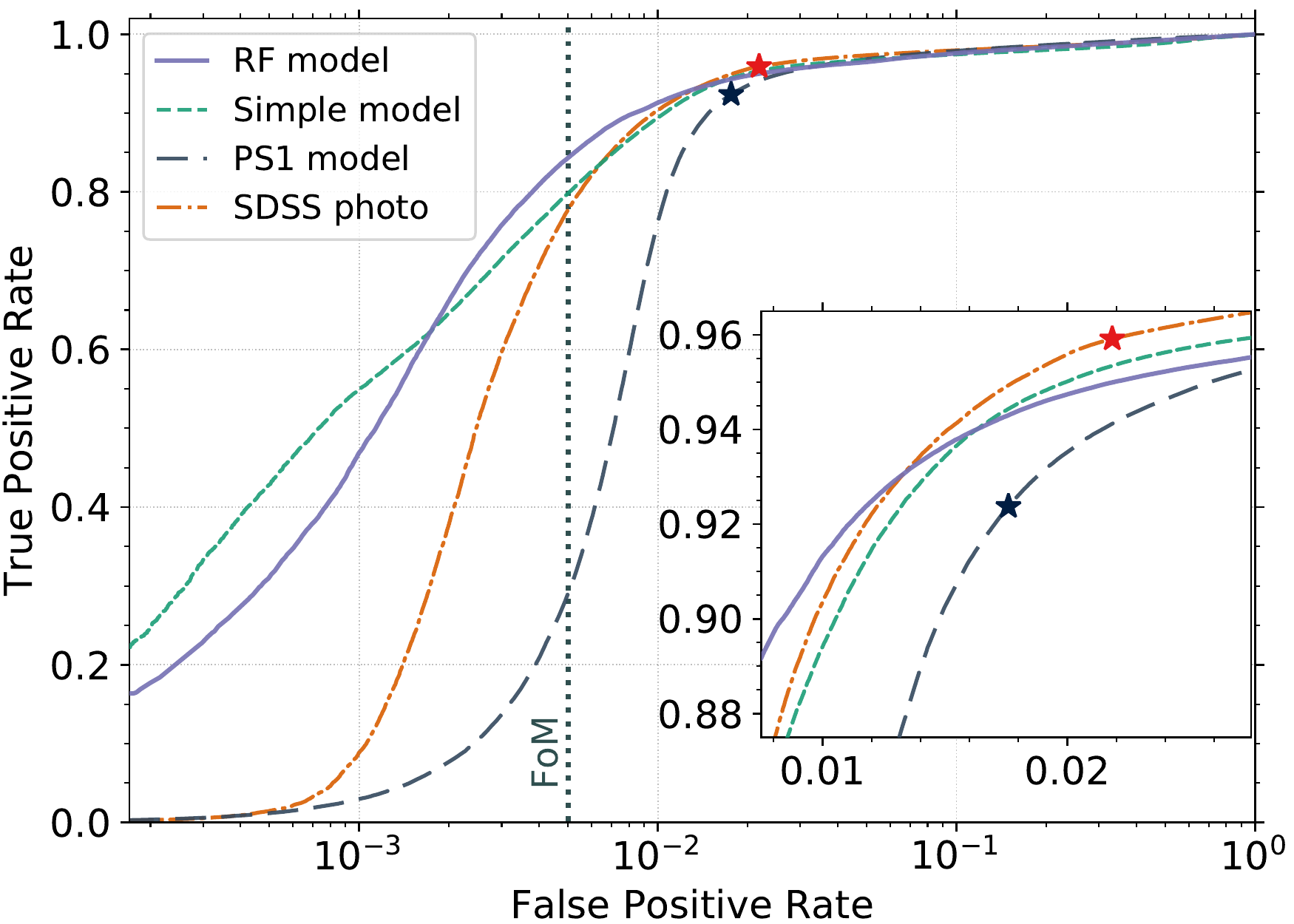}
  \caption{ROC curves comparing the relative performance of the SDSS (orange
  dot-dashed line), PS1 (slate grey dashed line), simple (green dotted
  line), and RF (solid purple) models as tested by the SDSS test set. Note
  that the FPR is shown on a logarithmic scale. The vertical dotted line
  shows $\mathrm{FPR} = 0.005$, corresponding to the FoM. The inset shows a
  zoom in around the region where the ROC curves cross (see text for further
  details). The black and red stars show the FPR and TPR if adopting the PS1
  model and SDSS \textit{photo} classification cuts, respectively. The RF
  model delivers the highest FoM.}
  \label{fig:roc_sdss}
\end{figure}

While CV on the \textit{HST} training set shows that the RF model
outperforms the alternatives, here, we test each of the previous models with
the SDSS training set, which provides an independent set of $\sim$3.8$\times
10^6$ sources with high-confidence labels. Additionally, the use of SDSS
spectra allows us to compare our new models to the classifications from the
SDSS \texttt{photo} pipeline, hereafter the SDSS model, which soundly
outperformed the PTF point source classification model \citep{Miller17}. We
create an ROC curve for the SDSS model by thresholding on the ratio of PSF
flux to \texttt{cmodel} flux measured in the SDSS images (see
\citealt{Miller17} for more details).

To compare the 4 models, we evaluate the performance of each model on the
subset of SDSS training set sources that have $i_\mathrm{PS1}$ detections
(to compare with the PS1 model) and SDSS \texttt{photo} classifications (to
compare with the SDSS model). We further exclude QSOs with $z < 1$ (=
133,856 sources; QSOs are typically considered point sources but low-$z$
QSOs can have resolved host galaxies; see \citealt{Miller17}), and galaxies
with $z < 10^{-4}$ (= 13,261 sources; such low $z$ is only expected in the
local group meaning most of these classifications are likely spurious). In
total, this subsample (hereafter the SDSS test set) includes 3,592,940
sources from the SDSS training set.

ROC curves for the RF, simple, PS1, and SDSS models, as measured by the SDSS
test set, are shown in Figure~\ref{fig:roc_sdss}. The FoM for the PS1,
simple, and RF models is higher as tested by SDSS spectroscopic sources
because the SDSS training set contains brighter, higher SNR (and hence
easier to classify) sources. As before, we find that the FoM for the RF
model is superior to the alternatives. Interestingly, we also find that the
ROC curves cross, and that the SDSS model provides the largest TPR for
$\mathrm{FPR} \gtrsim 0.015$. That the RF and SDSS curves cross suggests
that there may be regimes where the SDSS \texttt{photo} classifications are
superior to the RF model. Below, we argue that a bias in the SDSS training
set is amplified by a bias in the SDSS \texttt{photo} classification, which
is why these curves cross.

\begin{figure*}[htb]
 \centering
  \includegraphics[width=6.2in]{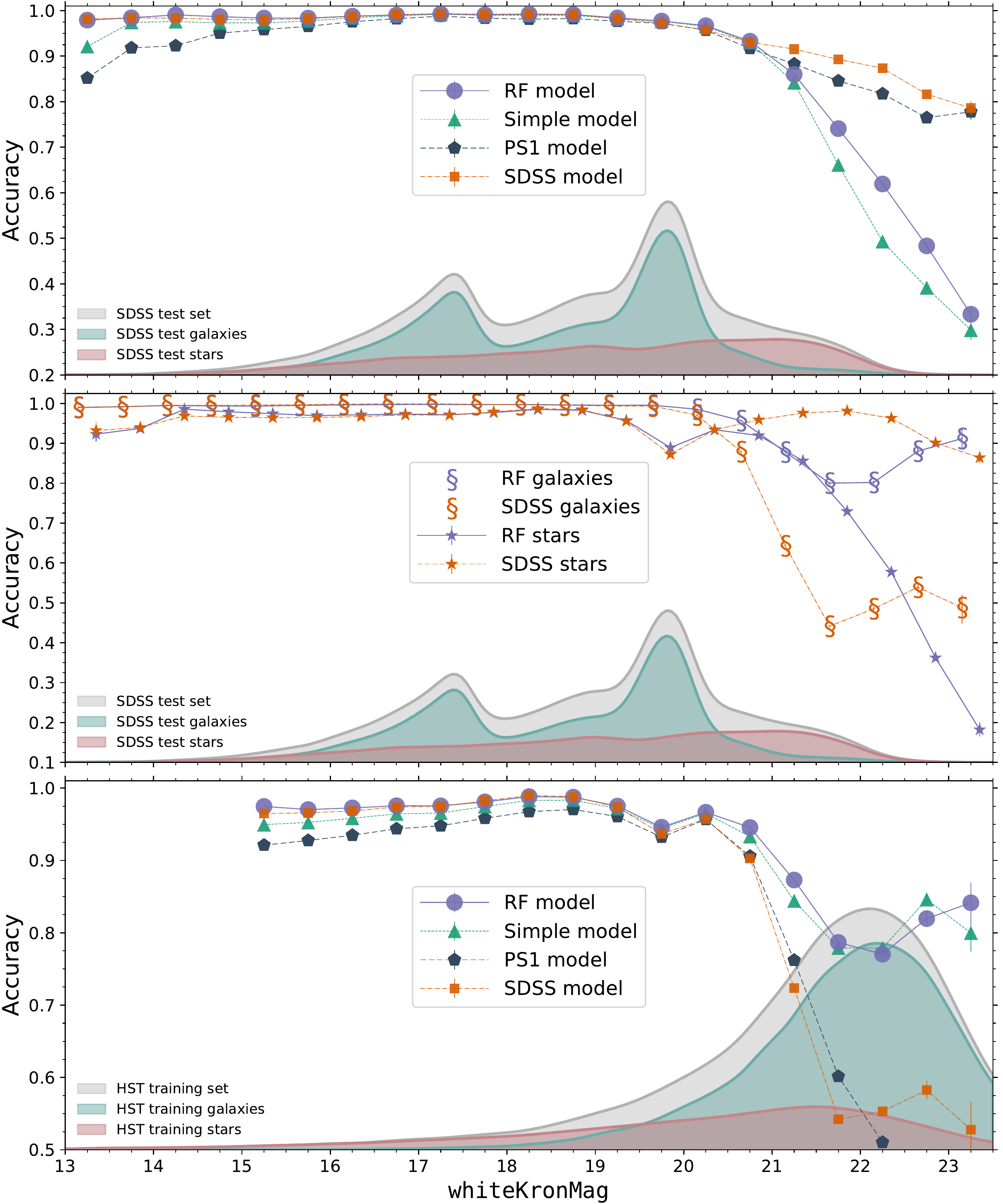}
  \caption{ Model accuracy for the RF (purple circles), simple (green
  triangles), PS1 (slate gray pentagons), and SDSS (orange squares) models
  as a function of \texttt{whiteKronMag} evaluated on the SDSS test set. The
  bin widths are 0.5\,mag, and the error bars represent the central 68\%
  interval from bootstrap resampling within each bin.
  \textit{Top}: Model accuracy curves for the SDSS test set. This panel also
  shows a Gaussian KDE of the PDF for the SDSS test set, as well as the
  point sources (labeled stars) and extended objects (labeled galaxies) in
  the SDSS test set in the shaded gray, red, and green regions,
  respectively. The amplitude of the point source and extended object PDFs
  have been normalized by their relative fraction of the full test set.
  \textit{Middle}: SDSS test set accuracy curves for individual point
  sources and extended objects, equivalent to the TPR and TNR, respectively,
  as classified by the RF and SDSS models (the simple and PS1 models are not
  shown for clarity). Note -- all 3 panels have the same bin centers, though
  here the markers are slightly offset for clarity. The SDSS model
  classifies faint point sources correctly, but has poor performance on
  faint extended objects, while the opposite is true for the RF model.
  \textit{Bottom}: The accuracy curves for all 4 models following the
  bootstrap procedure (described in the text) to correct for the SDSS test
  set bias whereby point sources outnumber extended objects at
  $\mathtt{whiteKronMag} \gtrsim 20.5\,\mathrm{mag}$. The PDFs shown in this
  panel are derived from KDEs of the \textit{HST} training set (as in
  Figure~\ref{fig:cvacc_hst}). After correcting for the SDSS test set number
  count bias, the RF and simple models produce more accurate classifications
  of faint sources than the SDSS and PS1 models.}
  \label{fig:acc_sdss}
\end{figure*}

Accuracy curves for each of the 4 models, as evaluated on the SDSS test set,
is shown in the top panel of Figure~\ref{fig:acc_sdss}. The RF, simple, and
SDSS models all provide near-perfect ($\ge 97.5$\%) accuracy down to
\texttt{whiteKronMag} $\approx 20\,\mathrm{mag}$. The PS1 model is similar,
though has a noticably worse performance for the brightest
(\texttt{whiteKronMag} $\lesssim 14.5\,\mathrm{mag}$) sources. For sources
with $\mathtt{whiteKronMag} > 21\,\mathrm{mag}$, the SDSS and PS1 models
provide far more accurate classifications than the RF and simple models. In
this faint regime, the SDSS test set is dominated by point sources (top
panel, Figure~\ref{fig:acc_sdss}). This is counter to what is observed in
nature (at high galactic latitudes), as extended object number counts exceed
those of point sources around $r \gtrsim 20\,\mathrm{mag}$ (e.g.,
\citealt{Yasuda01,Shanks15}). This bias in the SDSS test set is due to the
SDSS targeting proclivity for luminous red galaxies (LRGs) at $z \approx
0.5$ (e.g., \citealt{Eisenstein01}) and faint $z \approx 2.7$ QSOs (e.g.,
\citealt{Ross12}).\footnote{For the SDSS test set, the peak in the extended
object PDF at $\mathtt{whiteKronMag} \approx 19.75\,\mathrm{mag}$ is
dominated by LRGs, while the population of faint ($\mathtt{whiteKronMag} >
21\,\mathrm{mag}$) point sources is dominated by QSOs.}

In addition to this bias in the SDSS test set, the SDSS (and PS1) model are
biased towards classifying faint resolved sources as unresolved. This is due
to the hard cut on a single value of the PSF to \texttt{cModel} (or
\texttt{Kron} for PS1) flux ratio. The reason for this can easily be seen in
the top left panel of Figure~\ref{fig:features}, where a classification cut
at $\mathtt{whitePSFKronRatio} = 0.875$ (equivalent to the SDSS cut)
correctly identifies nearly all of the point sources, but does a
particularly bad job with the faintest extended objects. At low SNR the
large scatter in flux ratio measurements results in many misclassifications.

We show further evidence for this classification bias in the middle panel of
Figure~\ref{fig:acc_sdss}, which shows the accuracy with which individual
extended sources and point sources in the SDSS test set are
classified\footnote{This is equivalent to showing the true negative rate
(TNR = TN/[TN + FP]) and TPR, respectively.} by the RF and SDSS models
(curves for the simple and PS1 models show similar trends as the RF and SDSS
models, respectively, but are omitted for clarity). For faint
($\mathtt{whitePSFKronRatio} > 21\,\mathrm{mag}$) sources, the SDSS model
performs well on point sources ($\mathrm{TPR} \gtrsim 0.9$) and poorly on
extended sources ($\mathrm{TNR} \approx 0.5$). The opposite is true for the
RF model, with $\mathrm{TNR} \gtrsim 0.8$ and a TPR that declines to
$\sim$0.2 for the faintest SDSS test set sources. Thus, for faint sources
the RF model is slightly biased towards resolved object classifications,
however, this bias is in line with what is observed in nature. These
classification biases, taken together with the SDSS test set bias towards
point sources at the faint end, explain why the accuracy curves for the SDSS
(and PS1) model outperform the RF (and simple) model (and also why their ROC
curves cross in Figure~\ref{fig:roc_sdss}).

The bottom panel of Figure~\ref{fig:acc_sdss} shows that after correcting
for the number counts bias in the SDSS test set, the RF and simple models
greatly outperform the SDSS and PS1 models. We correct for the number count
bias via bootstrap resampling, whereby we select a subset of point sources
and extended objects from the SDSS test set to match the ratio of point
sources to extended objects in the \textit{HST} training set. The
\textit{HST} training set, which is selected photometrically, should serve
as a far better approximation for the relative number counts of point
sources and extended objects at high-Galactic latitudes than the SDSS test
set. The bootstrap occurs in bins of width 0.5\,mag from
\texttt{whiteKronmag} = 15\,mag to 23.5\,mag, and we select 100 bootstrap
samples within each bin. In each bin the size of the bootstrap sample is set
by the underrepresented class within the SDSS test set. For example, if the
\textit{HST} training set has an unresolved--resolved number ratio of 0.6
and in the same bin the SDSS test set has 1000 point sources and 4000
extended objects, then 1000 point sources and 1667 extended objects will be
selected in each bootstrap sample. Similarly, for an unresolved--resolved
number ratio of 0.25 in a bin with 800 point sources and 1000 extended
objects, then 250 point sources and 1000 extended objects will be selected.

Correcting for the number counts bias in the SDSS test set reveals some
interesting trends: as was the case prior to correction all 4 models perform
similarly well for bright (\texttt{whiteKronMag} $\lesssim
20\,\mathrm{mag}$) sources. However, for fainter sources the RF and simple
models significantly outperform the SDSS and PS1 models. The bottom panel of
Figure~\ref{fig:acc_sdss} also shows a kink at \texttt{whiteKronMag}
$\approx 19.75\,\mathrm{mag}$. As first explained in \citet{Miller17}, this
kink is due to blended, faint red stars that were targeted as candidate
LRGs. Thus, spectra show these sources to be stellar, while they appear
extended in imaging data. Finally, we conclude that for source distributions
similar to what is observed in nature, the RF model outperforms the
alternatives discussed here in both the FoM and the overall accuracy.


\section{The PS1 Catalog Deployed: Integration in ZTF}
\label{sec:ztf}

\subsection{The Zwicky Transient Facility}

While we have developed a general model to identify point sources, the
resulting RF classifications have been specifically deployed in the Zwicky
Transient Facility\footnote{\url{http://www.ztf.caltech.edu/}} (ZTF;
\citealt{Bellm:18:ZTF, Dekany:18:ZTF}) real-time pipeline
\citep{Masci:18:ZTF}. Briefly, ZTF is the next-generation Palomar
time-domain survey, which succeeds PTF \citep{Rau09, Law09} and the
intermediate Palomar Transient Factory (iPTF; \citealt{Kulkarni13}). ZTF,
with its 47\,deg$^2$ field of view, can scan at a rate $\sim$15$\times$
faster than PTF/iPTF ($>3{,}750\,\deg^2\,\mathrm{hr}^{-1}$) to a depth of
$R_\mathrm{ZTF} \approx 20.4$\,mag ($5\sigma$). ZTF will observe the entire
sky with $\delta > -30^{\circ}$ $\sim$300 times per year, with publicly
distributed alerts on newly observed positional or flux variability released
in near real time \citep{Patterson:18:ZTF}.\footnote{See: \url{https://ztf.uw.edu/alerts/public/} for real-time alerts.}

\subsection{Integration in the ZTF Alert Stream}

An initial, and pressing, question for filtering the ZTF alert stream, is:
does the newly identified variable have a Galactic or extragalactic origin?
Hence the need for a resolved--unresolved model, and in particular, one that
is deeper than typical ZTF observations (to identify faint stars flaring
above the ZTF detection limit). While ZTF will address many science
objectives (e.g., \citealt{Graham:18:ZTF}), a primary motivation is the
search for fast transients, especially kilonovae (KNe), the result of
merging binary neutron stars. If the proximity and sky location of a KN is
favorable, these events can be detected via gravitational waves (e.g.,
GW\,170817, see \citealt{Abbott17} and references therein). The search for
KNe is plagued by significant foreground contamination in the form of
stellar flares and/or orbital modulation (e.g., \citealt{Kulkarni06,
Berger12, Kasliwal16}). Our PS1 RF model enables the systematic removal of
faint stars from extragalactic candidate lists, and our adopted FoM ensures
that nearly every galaxy ($\sim$99.5\%) is searched for candidate KNe.

Newly discovered ZTF candidates are associated with the 3 nearest PS1
counterparts within 30\arcsec\ in the real-time alert packets
\citep{Masci:18:ZTF}. Counterparts are selected from ZTF calibration
sources, which includes all PS1 \textit{MeanObject} table sources with
$\mathtt{nDetections} \ge 3$. Thus, to create the ZTF--PS1 point source
catalog we selected sources from the \textit{StackObjectAttributes} table
with $\mathtt{nDetections} \ge 3$, and merged these classifications with the
ZTF calibration sources. Ultimately, non-unique sources (i.e., if a single
\texttt{objID} corresponds to multiple rows with \texttt{primaryDetection} =
1) are excluded from the classification catalog.

\begin{figure}[htb]
 \centering
  \includegraphics[width=3.35in]{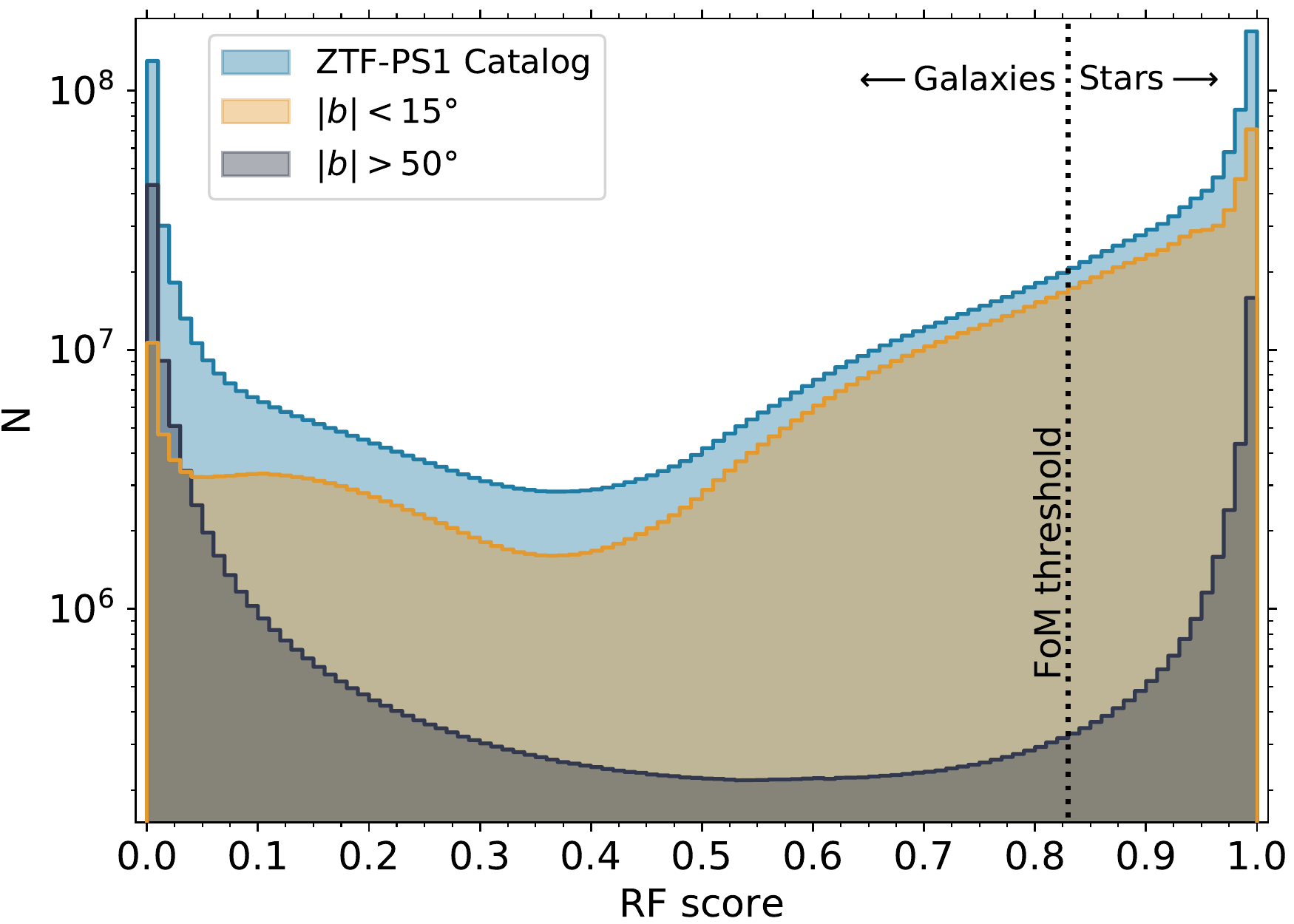}
  \caption{ The distribution of RF classification score for all sources in
  the ZTF--PS1 point source catalog. Note that the number counts are shown on
  a log scale. The vertical dotted line shows the FoM-optimized
  classification threshold, sources to the right of the line are classified
  as point sources. The full catalog is shown in blue, while Galactic plane
  sources ($|b| < 15^{\circ}$) are shown in orange, and high galactic
  latitude sources ($|b| > 50^{\circ}$) are shown in grey. Ambiguous
  classifications ($0.2 \lesssim \mathrm{RF\;score} \lesssim 0.8$) in the
  catalog are dominated by sources in the Galactic plane.}
  \label{fig:ztf_hist}
\end{figure}

In total, there are 1,484,281,394 PS1 sources with RF
classifications.\footnote{An additional 8,520,167 sources with
significant parallax or proper motion have $\mathrm{RF\;score} = 1$ in the
ZTF database (see \S\ref{sec:gaia}).} A histogram showing the distribution
of the final RF classifications is shown in Figure~\ref{fig:ztf_hist}. The
thresholds appropriate for identifying point-source counterparts to the ZTF
candidates are reported in Table~\ref{tbl:fpr} (note that these thresholds
apply to $\mathtt{nDetections} \ge 3$ sources). Of the $\sim$1.5$\times
10^{9}$ sources in the ZTF--PS1 catalog, 734,476,355 ($\sim$50\%) are
classified as point sources using the FoM-optimized classification threshold of 0.83.

\begin{deluxetable*}{lcc|lccccc}
    \tablecolumns{9} 
    \tablewidth{0pt} 
    \tablecaption{Classification thresholds for the ZTF--PS1 Catalog.
\label{tbl:fpr}}
    \tablehead{ 
    \colhead{Selection criteria} & \colhead{$N$\tablenotemark{a}} & \colhead{Accuracy\tablenotemark{b}} & \colhead{FPR} & \colhead{0.005} & \colhead{0.01} & \colhead{0.02} & \colhead{0.05} & \colhead{0.1}
    }
    \startdata
\multirow{2}{*}{All sources} & \multirow{2}{*}{35,007} & \multirow{2}{*}{$93.9\pm0.1$\%} &                     TPR  & $0.734\,^{+0.012}_{-0.014}$ & $0.792\,^{+0.010}_{-0.009}$ & $0.843\,^{+0.008}_{-0.008}$ & $0.904\,^{+0.005}_{-0.005}$ & $0.947\,^{+0.004}_{-0.004}$  \\
\multicolumn{1}{l}{}                 & & & Threshold & $0.829\,^{+0.018}_{-0.010}$ & $0.724\,^{+0.016}_{-0.014}$ & $0.597\,^{+0.014}_{-0.010}$ & $0.397\,^{+0.008}_{-0.006}$ & $0.224\,^{+0.006}_{-0.004}$ 
\\ \hline
\multirow{2}{*}{$\mathtt{rKronMag} < 21$} & \multirow{2}{*}{13,570} & \multirow{2}{*}{$98.0\pm0.1$\%} &
                                              TPR  & $0.797\,^{+0.146}_{-0.101}$ & $0.964\,^{+0.007}_{-0.014}$ & $0.980\,^{+0.003}_{-0.004}$ & $0.989\,^{+0.003}_{-0.002}$ & $0.995\,^{+0.002}_{-0.001}$  \\
                                                & & & Threshold & $0.970\,^{+0.015}_{-0.043}$ & $0.645\,^{+0.115}_{-0.041}$ & $0.406\,^{+0.048}_{-0.014}$ & $0.170\,^{+0.018}_{-0.011}$ & $0.069\,^{+0.006}_{-0.003}$     \\ \hline
\multirow{2}{*}{$\mathtt{rKronMag} < 20$}  & \multirow{2}{*}{6,956}& \multirow{2}{*}{$99.0\pm0.1$\%} &
                                              TPR  & $0.697\,^{+0.197}_{-0.219}$ & $0.954\,^{+0.038}_{-0.082}$ & $0.994\,^{+0.002}_{-0.002}$ & $0.997\,^{+0.002}_{-0.002}$ & $0.998\,^{+0.001}_{-0.001}$   \\
                                                 & &  & Threshold & $0.993\,^{+0.005}_{-0.015}$ & $0.923\,^{+0.052}_{-0.328}$ & $0.339\,^{+0.082}_{-0.050}$ & $0.132\,^{+0.036}_{-0.021}$ & $0.047\,^{+0.006}_{-0.005}$
    \enddata
    \tablecomments{10-fold CV is performed on the entire \textit{HST} training set, but the metrics reported here include only sources that satisfy the selection criteria defined by the first column and $\mathtt{nDetections} \ge 3$. The reported uncertainties represent the central 90\% interval from 100 bootstrap resamples of the training set.} 
    \tablenotetext{a}{Number of \textit{HST} training set sources within the selected subset.}
    \tablenotetext{b}{Classification accuracies are reported relative to a $\mathrm{RF \; score} = 0.5$ classification threshold.}
\end{deluxetable*}

Figure~\ref{fig:ztf_hist} additionally shows that most of the point sources
in the catalog are located in the Galactic plane, most of the
(high-confidence) extended objects are outside the plane, and
(unsurprisingly) that classification is more challenging in regions of high
stellar density. At high galactic latitudes ($|b| > 50^{\circ}$), where the
distribution of sources is similar to the \textit{HST} training set, sources
are well segregated (RF score $\approx$0 or 1), with very few ambiguous
classifications ($0.2 \lesssim \mathrm{RF\;score} \lesssim 0.8$). The
Galactic plane region ($|b| < 15^{\circ}$) dominates the ambiguous
classifications in the ZTF--PS1 point source catalog. We attribute this to a
lack of reliable training data in high-stellar-density regions, and
significantly more blending, which results in point sources appearing
extended. Thus, identifying stellar sources in the Galactic plane likely
requires a lower threshold than the FoM-optimized classification value. The
final tuning of the resolved--unresolved classification thresholds is a
critical early step in the filtering of ZTF candidates (e.g.,
\citealt{Kasliwal:18:ZTF}), which is necessary to optimize follow-up of
newly discovered transients.

\subsection{Verifying, and Updating, the Catalog with \textit{Gaia}}
\label{sec:gaia}

The \textit{Gaia} satellite \citep{Gaia-Collaboration16} is currently
conducting an all-sky survey that provides unprecedented astrometric
accuracy in measuring the positions, parallaxes, and proper motions of
$\sim$1.3 billion sources (\citealt{Gaia-Collaboration18}; an additional
$\sim$0.3 billion sources have just position measurements). The
\textit{Gaia} selection function is biased against resolved galaxies
\citep{Gaia-Collaboration16}, so it cannot provide a symmetric test of the
ZTF--PS1 catalog. Nevertheless, \textit{Gaia} has identified hundreds of
millions of stars that can be used to test our classifications.

Given that \textit{Gaia} does not classify the sources it detects as either
resolved or unresolved, we test the ZTF--PS1 catalog by selecting a pure
sample of \textit{Gaia} stars based on high-significance parallax, $\varpi$,
and proper motion, $\mu$, measurements.\footnote{Given the large distances,
\textit{Gaia} will measure low SNR $\varpi$ and $\mu$ for extragalactic
(i.e., extended) sources.} We define the $\varpi$ significance as
$\varpi$/$\sigma_\varpi$ (called \texttt{parallax\_over\_error} in the
\textit{Gaia} database). We obtain the total proper motion $\mu$ by adding
the proper motion in Right Ascension $\mu_{\alpha\ast}$ and Declination
$\mu_{\delta}$ (\texttt{pmra} and \texttt{pmdec} in the database,
respectively) in quadrature. The uncertainty in the total proper motion,
$\sigma_\mu$, is calculated via the proper motion uncertainties in Right
Ascension, $\sigma_{\mu_{\alpha\ast}}$, and Declination,
$\sigma_{\mu_{\delta}}$, while accounting for the correlation coefficient
between $\mu_{\alpha\ast}$ and $\mu_{\delta}$,
$\rho(\mu_{\alpha\ast}\mu_{\delta})$:

$$ \sigma_\mu^2 = \frac{\mu_{\alpha\ast}^2}{\mu^2}\sigma_{\mu_{\alpha\ast}}^2 +
\frac{\mu_{\delta}^2}{\mu^2}\sigma_{\mu_{\delta}}^2 +
2\frac{\mu_{\alpha\ast}\mu_{\delta}}{\mu^2} \rho(\mu_{\alpha\ast}\mu_{\delta})
\sigma_{\mu_{\alpha\ast}}\sigma_{\mu_{\delta}}.$$
The proper motion signficance is then defined as $\mu/\sigma_\mu$.

To determine the threshold for ``high significance'' in $\varpi$ and $\mu$
we use faint stars and activate galactic nuclei (AGN), respectively. In
\citet{Lindegren18}, a sample of $\sim$5.5$\times10^{5}$ AGN with
\textit{Gaia} observations were identified, and the cosmological distances
to these sources mean they should have extremely small parallaxes and proper
motions. We find that 99.997\% of these AGN have $\mu$ significance $<
5.62$, and the highest $\mu$ significance in the entire AGN sample is
$\sim$7.42. Thus, we apply a conservative cut of 7.5 on $\mu$ significance
to select a pure set of stars with little contamination from extended,
extragalactic objects.

The AGN sample is not sufficient for defining a cut on $\varpi$ significance,
because \citet{Lindegren18} require $\varpi$/$\sigma_\varpi < 5$. Instead,
we use the 20,568,254 \textit{Gaia} sources with $\varpi$ and $\mu$
measurments, $20.5\,\mathrm{mag} \le G \le 21\,\mathrm{mag}$, where $G$ is
the mean brightness measured by \textit{Gaia} in the $G$ filter, and that
pass the cuts defined by Eqn.~(C.1) and (C.2) in \citet{Lindegren18}. These
cuts are designed to remove low-confidence parallax measurements in regions
of high stellar density. The low SNR detections for these \textit{Gaia}
sources provide an estimate of the scatter of the parallax significance, as
they are too faint for high-significance detections.\footnote{The typical
uncertainty on parallax for sources this faint is $\sigma_\varpi \approx
2$\,mas \citep{Lindegren18}. While some of these $\sim$2$\times10^7$ sources
may be at a distance $<$500\,pc away, that will not be true for the vast
majority, and thus they will not have significant parallax measurements.}
Looking at the full distribution of parallax significance for these faint
sources we find that 99.997\% have parallax significance $< 7.94$. Again, we
apply a conservative cut of parallax significance $\ge$8 to select bonafide
stars from the \textit{Gaia} data.

\begin{figure}[htb]
 \centering
  \includegraphics[width=3.3in]{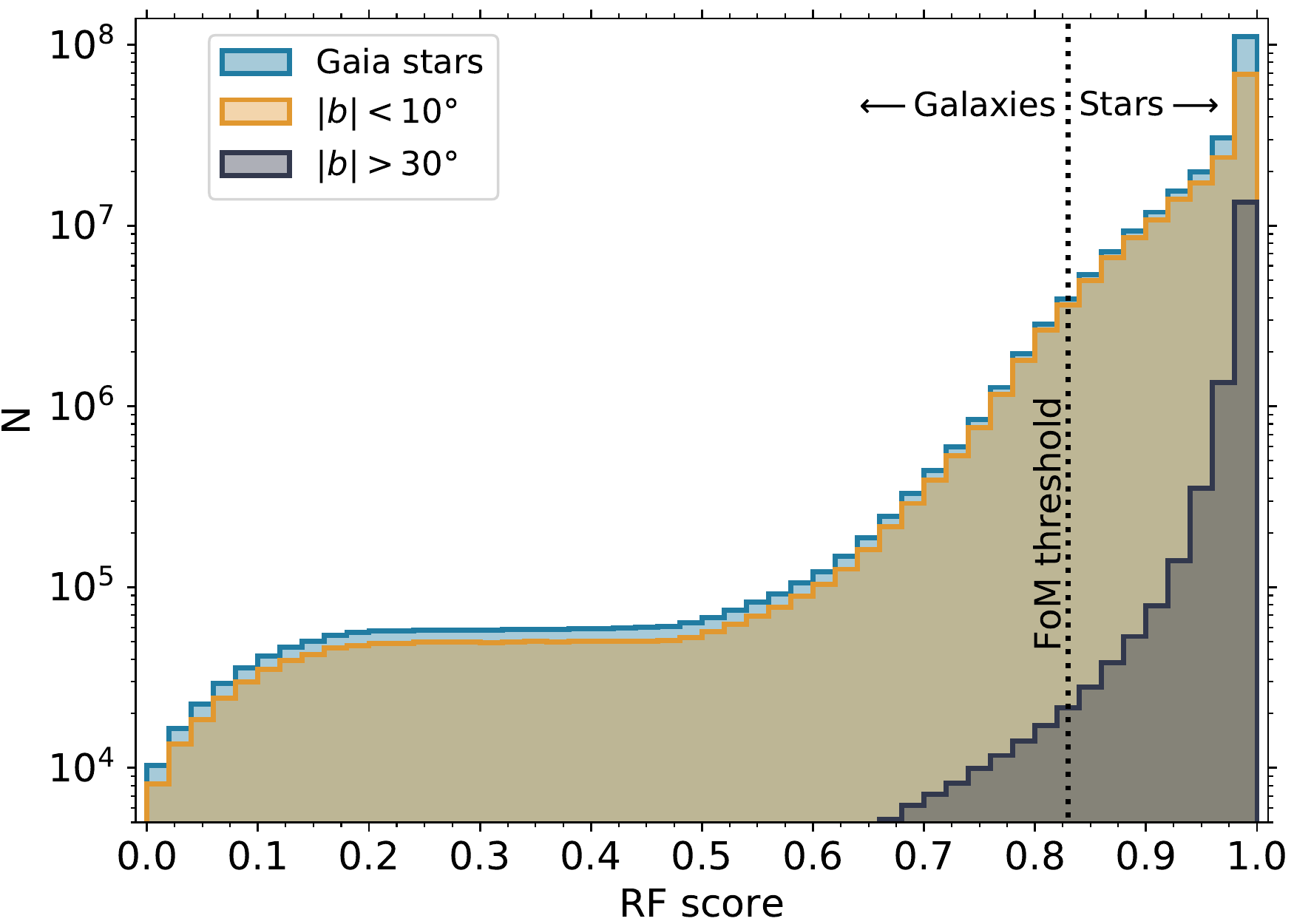}
  \caption{ 
  The distribution of RF classification score for sources in the ZTF--PS1
  star--galaxy catalog selected as high-probability stars from \textit{Gaia}
  due to their significant proper motion (see text for further details).
  Note that the number counts are shown on a log scale. The vertical dotted
  line shows the FoM-optimized classification threshold, sources to the
  right of the line are classified as stars. The full catalog is shown in
  blue, while Galactic plane sources ($|b| < 10^{\circ}$) are shown in
  orange, and high galactic latitude sources ($|b| > 30^{\circ}$) are shown
  in grey. Less than 0.25\% of these stars have $\mathrm{RF\;score} \lesssim
  0.5$.
  }
  \label{fig:gaia}
\end{figure}

Using the existing crossmatch between \textit{Gaia} and PS1,\footnote{See
\citet{Marrese17} for details on \textit{Gaia} crossmatching external
catalogs. There are 810,359,898 \textit{Gaia} sources crossmatched with PS1,
see:
\url{https://gea.esac.esa.int/archive/documentation//GDR2/Catalogue_consolidation/chap_cu9val_cu9val/ssec_cu9xma/sssec_cu9xma_extcat.html}
.} we have identified 38,764,553 and 234,176,264 high-confidence stars that
pass the cuts defined by Eqn.~(C.1) and (C.2) in \citet{Lindegren18} and
have parallax significance $\ge 8$ or proper motion significance $\ge 7.5$,
respectively. Of these, 35,599,830 and 225,682,755 have respective
counterparts in the ZTF--PS1 catalog (the respective differences of
3,164,723 and 8,493,509 correspond to sources with either 0 or $>$1 entries
in the PS1 \textit{StackObjectThin} table). For the proper motion selected
stars, we show the distribution of RF classification scores for these
stellar objects in Figure~\ref{fig:gaia}. It is clear from
Figure~\ref{fig:gaia} that the vast majority of stars are classified
correctly. Half of the stars selected via proper motion have
$\mathrm{RF\;score}\ge 0.99$, while 98.1\% have $\mathrm{RF\;score}\ge
0.83$, the FoM classification threshold, and 99.75\% have
$\mathrm{RF\;score}\ge 0.5$, the traditional binary classification
threshold. The percentages are even higher for the parallax-selected sample.
These calssification results are significantly better than those reported in
Table~\ref{tbl:fpr}, which makes sense given that this high significance
\textit{Gaia} sample is much brighter than the sources in the \textit{HST}
training set (median brightness $G = 18.0\,\mathrm{mag}$, 95$^\mathrm{th}$
percentile brightness $G = 19.7\,\mathrm{mag}$). Nevertheless, we conclude
that the \textit{Gaia} data confirms that our method does an excellent job
of identifying point sources.

Finally, the 8,520,167 sources selected by either the parallax or proper
motion cuts described above that do not have counterparts in the ZTF--PS1
catalog are assigned an RF classification score of 1 in the ZTF database.
Thus, new transient candidates with positions consistent with these sources
will be flagged as likely stars.

\section{Summary and Conclusions}

We have presented the development of a large ($\sim$1.5$\times 10^{9}$),
deep ($m \lesssim 23.5\,\mathrm{mag}$) catalog of point sources and extended
objects based on PS1 data. We classify these sources using a machine
learning framework built on a RF model. The RF model is trained using 47,093
PS1 sources with \textit{HST} COSMOS morphological classifications.

To construct the RF model, we introduced ``white flux'' features, which
correspond to a weighted mean of the relevant features over the
$grizy_{\mathrm {PS1}}$ filters in which a source is detected. The ``white
flux'' features allow us to classify all PS1 sources, irrespective of the
filters in which the source was detected or the line-of-sight reddenning.
One of these newly created features, \texttt{whitePSFKronDist}, is useful on
its own for separating stars and galaxies. Unlike a hard cut on the PSF and
Kron flux ratio, as is employed by the SDSS and PS1 models,
\texttt{whitePSFKronDist} retains knowledge of the SNR and therefore can
provide higher confidence classifications. From \texttt{whitePSFKronDist} we
created the simple model, which does a good job of separating point sources
and extended objects. Ultimately, the 11 ``white flux'' features, used in
combination with the RF algorithm, provide the best classification of PS1
sources.

CV on the \textit{HST} training set shows that the RF (FoM$ = 0.71$) and
simple (FoM$ = 0.657$) models greatly outperform the PS1 (FoM$ = 0.007$)
model. For faint sources ($\mathtt{whiteKronMag} > 20$\,mag) the PS1 model
misclassifies many extended objects as point sources, while both the simple
and RF models provide overall classification accuracies $\gtrsim 85$\% as
faint as $\mathtt{whiteKronMag} = 23$\,mag.

We find that when evaluated with the SDSS test set, the SDSS and PS1 models
provide more accurate classifications than the RF and simple models,
especially for faint ($\mathtt{whiteKronMag} \gtrsim 21$\,mag) sources. This
reversal, relative to the \textit{HST} training set results, can be
attributed to a bias in the SDSS test set and the SDSS classification model.
In the SDSS test set point sources outnumber galaxies at the faint end,
which is counter to what is observed (at high galactic latitudes).
Furthermore, the SDSS and PS1 models, which utilize a hard cut on flux
ratios, are likely to classify low SNR sources as point sources. Together,
these effects amplify the perceived performance of the SDSS and PS1 models.
Using a bootstrap resampling procedure, we correct for the relative number
counts bias in the SDSS test set, and find that the RF and simple models
outperform the SDSS and PS1 models, both in terms of FoM and overall
accuracy. Thus, of the 4 models considered in this study the RF model is
superior to all others.

We have deployed the RF model in support of the ZTF real-time pipeline,
resulting in the classification of $\sim$1.5$\times 10^{9}$ sources. The
catalog is dominated by point sources in the vicinity of the Galactic plane,
though we find that there are more extended objects than point sources at
high galactic latitudes, as is expected at the depth of PS1. ZTF is
currently producing public alerts for newly discovered variability, and the
ZTF--PS1 catalog is essential for removing the numerous foreground of
stellar flares, false positives in the search for fast transients and KNe,
from the extragalactic alert stream. The final ZTF--PS1 catalog is available
at MAST via
\dataset[doi:10.17909/t9-xjrf-7g34]{http://dx.doi.org/10.17909/t9-xjrf-7g34}.

Moving forward, future data releases and additional scrutiny of
\textit{Gaia} data will significantly increase the fidelity of the PS1
resolved--unresolved classification model. The \textit{HST} training set has
very few bright sources and no sources at low Galactic latitudes, leading to
less confident classifications in these regions (see
Figure~\ref{fig:ztf_hist}). As a space-based observatory, \textit{Gaia} will
resolve many stellar blends in the Galactic plane and identify millions of
stars brighter than 16\,mag. Many of the ambiguous classifications in the
ZTF--PS1 catalog (see \S\ref{sec:ztf}) will be directly identified as stars
due to their high proper motions and parallaxes (similar to the analysis in
\ref{sec:gaia}, though future \textit{Gaia} observations will lead to even
better precision). As previously noted, \textit{Gaia} does not downlink
measurements for extended sources. Thus, while \textit{Gaia} would allow us
to increase the size of our training set by many orders of magnitude, it
would also introduce a significant class imbalance. Correcting for the lack
of galaxies would require new approaches beyond those described here. It
should also be noted that \textit{Gaia} alone is not sufficient for our
purposes, as it only detects sources with $G \lesssim 21$\,mag, which does
not include the faint, flaring stars that we expect to be the primary false
positive in the search for fast transients. Nevertheless, our ability to now
merge several $\sim$all-sky surveys provides unprecedented power in the
classification of astronomical sources. This power is particularly important
for improving the scientific output and follow-up efficiency of time-domain
surveys.

\acknowledgements

This work would not have been possible without the public release of the
PS1, SDSS, and \textit{Gaia} data. We are particularly grateful to the MAST
PS1 team for answering several inquiries regarding the PS1 data, and
especially B.~Shiao, who helped us navigate the PS1 database. We thank
M.~Graham and A.~Mahabal for early conversations regarding the training of
the RF model, and B.~Bue for a discussion about cross validation strategies.

Y.T.\ is funded by JSPS KAKENHI Grant Numbers JP16J05742. Y.T.\ studied as a
Global Relay of Observatories Watching Transients Happen (GROWTH) intern at
Caltech during the summer and fall of 2017. GROWTH is funded by the National
Science Foundation under Partnerships for International Research and
Education Grant No 1545949. A.A.M.\ is funded by the Large Synoptic Survey
Telescope Corporation in support of the Data Science Fellowship Program.

Based in part on software developed as a part of the Zwicky Transient Facility project, a scientific collaboration among the California Institute of Technology, the Oskar Klein Centre, the Weizmann Institute of Science, the Joint Space-Science Institute (via the University of Maryland, College Park), the University of Washington, Deutsches Elektronen-Synchrotron, the University of Wisconsin-Milwaukee, and the TANGO Program of the University System of Taiwan. Further support for ZTF is provided by the U.S. National Science Foundation under Grant No. AST-1440341.

\facility{PS1, Sloan}

\software{\texttt{astropy} \citep{Astropy-Collaboration13}, 
          \texttt{scipy} \citep{Jones01}, 
          \texttt{matplotlib} \citep{Hunter07},
          \texttt{pandas} \citep{McKinney10},
          \texttt{scikit-learn} \citep{Pedregosa12}}


\end{CJK*}
\end{document}